\documentclass[superscriptaddress,prb,twocolumn,showpacs]{revtex4}
\usepackage{amsmath}
\usepackage{amssymb}
\usepackage{graphicx}
\usepackage{bm}
\usepackage{dcolumn}

\begin{document}

\title{Bilayer splitting versus Fermi-surface warping as an origin of slow
oscillations of in-plane magnetoresistance in rare-earth tritellurides}
\author{P.D.~Grigoriev}
\affiliation{L. D. Landau Institute for Theoretical Physics, 142432 Chernogolovka, Russia}
\affiliation{National University of Science and Technology ``MISiS'', Moscow 119049,
Russia}
\affiliation{Institut Laue-Langevin, BP 156, 6 rue Jules Horowitz, 38042 Grenoble, France}
\author{A.A.~Sinchenko}
\affiliation{Kotel'nikov Institute of Radioengineering and Electronics of RAS, Mokhovaya
11-7, 125009 Moscow, Russia}
\affiliation{National Research Nuclear University (MEPhI), 115409 Moscow,Russia}
\affiliation{Univ. Grenoble Alpes, Inst. Neel, F-38042 Grenoble, France}
\affiliation{CNRS, Inst. Neel, F-38042 Grenoble, France}
\author{P.~Lejay}
\author{A.~Hadj-Azzem}
\author{J. Balay}
\author{O.~Leynaud}
\affiliation{Univ. Grenoble Alpes, Inst. Neel, F-38042 Grenoble, France}
\affiliation{CNRS, Inst. Neel, F-38042 Grenoble, France}
\author{V.N.~Zverev}
\affiliation{Institute of Solid State Physics, Chernogolovka, Moscow region 142432, Russia}
\affiliation{Moscow Institute of Physics and Technology, Dolgoprudnyi, Moscow region,
141700, Russia}
\author{P.~Monceau}
\affiliation{Univ. Grenoble Alpes, Inst. Neel, F-38042 Grenoble, France}
\affiliation{CNRS, Inst. Neel, F-38042 Grenoble, France}

\begin{abstract}
Slow oscillations (SlO) of the in-plane magnetoresistance with a frequency
less than 4 T are observed in the rare-earth tritellurides and proposed as
an effective tool to explore the electronic structure in various strongly
anisotropic quasi-two-dimensional compounds. Contrary to the usual
Shubnikov-de-Haas oscillations, SlO originate not from small Fermi-surface
pockets, but from the entanglement of close frequencies due to a finite
interlayer transfer integral, either between the two Te planes forming a
bilayer or between two adjacent bilayers. From the observed angular
dependence of the frequency and the phase of SlO we argue that they
originate from the bilayer splitting rather than from the Fermi-surface
warping. The SlO frequency gives the value of the interlayer transfer
integral $\approx 1$ meV for TbTe$_3$ and GdTe$_3$.
\end{abstract}

\date{\today }
\maketitle

\section{Introduction}

The measurement of magnetic quantum oscillations (MQO) and angular
magnetoresistance oscillations (AMRO) provides a powerful tool to study the
electronic properties of various quasi-two-dimensional (Q2D) layered
metallic compounds, such as organic metals (see, e.g., Refs. \cite%
{MQORev,OMRev,MarkReview2004,KartPeschReview} for reviews), cuprate and
iron-based high-temperature superconductors (see, e.g., \cite%
{HusseyNature2003,AbdelNature2006,ProustNature2007,AbdelPRL2007AMRO,McKenzie2007,DVignolle2008,HelmNd2009,HelmNd2010,BaFeAs2011,Graf2012}%
), heterostructures\cite{Kuraguchi2003}, graphite intercalation compounds,%
\cite{GraphiteIntercalatedNature2005} etc.

The Fermi surface (FS) of Q2D metals is a cylinder with weak warping $\sim
4t_{z}/E_{F}\ll 1$, where $t_{z}$ is the interlayer transfer integral and $%
E_{F}=\mu $ is the in-plane Fermi energy. The MQO with such FS have two
close fundamental frequencies $F_{0}\pm \Delta F$. In a magnetic field $%
\boldsymbol{B}=B_{z}$ perpendicular to the conducting layers $%
F_{0}/B=\mu/\hbar \omega_{c}$ and $\Delta F/B=2t_{z}/\hbar \omega _{c} $,
where $\hbar\omega_{c}={\hbar}eB_{z}/m^{\ast}c$ is the separation between
the Landau levels (LL), $m^{\ast}$ is an effective electron mass, and $c$
here is the light velocity.

The standard 3D theory of galvanomagnetic properties \cite%
{Abrik,Shoenberg,Ziman} is valid only at $t_{z}\gg \hbar \omega_{c}$, being
derived in the lowest order in the parameter $\hbar\omega _{c}/t_{z}$. This
theory predicts several peculiarities of magnetoresistance (MR) in Q2D
metals, such as AMRO \cite{KartsAMRO1988,Yam,Yagi1990} and the beats of MQO
amplitude.\cite{Shoenberg} One can extract the fine details of the FS, such
as its in-plane anisotropy \cite{Mark92} and its harmonic expansion \cite%
{Bergemann,GrigAMRO2010}, from the angular dependence of MQO frequencies and
from AMRO.

At $t_{z}\sim \hbar \omega _{c}$ several new qualitative features of MR
appear. At $\hbar \omega _{c}>t_{z}$ the strong monotonic growth of
longitudinal interlayer MR $R_{zz} (B_{z})$ was observed in various Q2D
metals \cite%
{SO,Coldea,PesotskiiJETP95,Zuo1999,W3,W4,Incoh2009,Wosnitza2002,Kang,WIPRB2012}
and explained recently \cite{WIPRB2012,WIPRB2011,WIPRB2013,GG2014}. At $%
t_{z}\gtrsim \hbar \omega _{c}$ the MR acquires the so-called slow
oscillations \cite{SO,Shub} and the phase shift of beats.\cite{PhSh,Shub}
These two effects are missed by the standard 3D theory \cite%
{Abrik,Shoenberg,Ziman} because they appear in the higher orders in $%
\hbar\omega_{c}/t_{z}$.

These slow oscillations (SlO) originate not from small FS pockets, but from
the finite interlayer hopping, because the product of oscillations with two
close frequencies $F_{0}\pm \Delta F$ gives oscillations with frequency $%
2\Delta F$. The conductivity, being a non-linear function of the oscillating
electronic density of states (DoS) and of the diffusion coefficient, has SlO
with frequency $2\,\Delta F\propto t_{z}$, while the magnetization, being a
linear functional of DoS, does not show SlO \cite{SO,Shub}. The SlO have
many interesting and useful features as compared to the fast quantum
oscillations. First, they survive at much higher temperature than MQO.
Second, they are not sensitive to a long-range disorder, which damps the
fast MQO similarly to finite temperature due to a spatial variation of the
Fermi energy. Therefore, the Dingle factor and the amplitude of SlO may be
much larger than those of usual MQO \cite{SO}. Third, the SlO allow to
measure the interlayer transfer integral $t_{z}$ and the in-plane Fermi
momentum $p_{F}\equiv \hbar k_{F}$. These features make the SlO to be a
useful tool to study the electronic properties of Q2D metals \cite{SO,Shub}.
Until now, the SlO were investigated only for the \textit{interlayer}
conductivity $\sigma _{zz}\left( B\right) $, when the current and the
magnetic field are both applied perpendicularly to the 2D layers, and only
in organic compounds \cite{SO,Shub,MarkReview2004}. At the same time, the
most of Q2D compounds, including pnictide high-temperature superconductors,
as a rule, have the shape of very thin flakes for which correct measurements
of the intralayer conductivity are reliable, especially in the case of good
metallic properties of studied compounds.

Very often the crystal consists of a stack of bilayers. In this case there
are two types of interlayer hopping integrals: the larger, $t_b$, is between
adjacent layers inside one bilayer, and the smaller one, $t_z$, is between
bilayers. Correspondingly, one may expect two types of SlO originating from
the bilayer and interbilayer electron hopping. The SlO from bilayer
splitting have not yet been studied.

Below we investigate the possibility and usefulness of SlO in the \textit{%
intralayer} electrical transport, choosing the non-organic layered Q2D
rare-earth tritelluride compounds $R$Te$_{3}$ ($R=$Y, La, Ce, Nd, Sm, Gd,
Tb, Ho, Dy, Er, Tm) as an example. Rare-earth tritellurides have an
orthorhombic structure ($Cmcm$) in the normal state and exhibit a $c$-axis
incommensurate charge-density wave (CDW) at high temperature, which was
recently a subject of intense studies \cite%
{Ru08,DiMasi95,Brouet08,SinchPRB12,Anis13,SSC14}. For the heaviest
rare-earth elements, a second $a$-axis CDW occurs at low temperature. In
addition to hosting incommensurate CDWs, magnetic rare-earth ions exhibit
closed-spaced magnetic phase transitions below 10 K \cite{Iyeri2003, Ru2008}
leading to coexistence and competition of many ordered states at low
temperatures. Therefore, any information about the Fermi surface on such
small energy scale beyond the ARPES resolution\cite{Brouet08,SchmittNJP2011}
is very important. An accurate measurement of $t_{z}$ as function of
temperature, provided by SlO, is also useful in these compounds. For the
possible observation of the SlO the rare-earth tritellurides are very
promising, because they have the appropriate anisotropy and good metallic
conductivity up to low temperatures. These compounds well illustrate our
goal: in addition to good metallic properties, their available single
crystals have a very flat shape, allowing correct measurements of the
intralayer conductivity \cite{Ru08}. Note that the RTe$_{3}$ compounds have
a doubled bilayer crystal structure, since there are two non-equivalent Te
bilayers in one elementary cell. Hence, this compound is a promising
candidate for the observation of SlO from bilayer splitting.

\section{Experiment}

For experiments we have chosen GdTe$_3$ and TbTe$_3$. Single crystals of
these compounds were grown by a self-flux technique under purified argon
atmosphere as described previously \cite{SinchPRB12}. Thin single crystal
samples with a thickness typically 0.1-0.3 $\mu$m were prepared by
micromechanical exfoliation of relatively thick crystals glued on a sapphire
substrate. The quality of selected crystals and the spatial arrangement of
crystallographic axes were controlled by X-ray diffraction. From
high-quality $[R(300 K)/R(10 K)>100]$ untwinned single crystals we cut
bridges with a length 200-500 $\mu$m and a width $50-80$ $\mu$m in well
defined, namely [100] and [001], orientations. Contacts for electrical
transport measurements in four-probe configuration have been prepared using
gold evaporation and cold soldering by In. The resistivity of the TbTe$_3$
samples typically $0.03$ m$\Omega$cm at room temperature was the same as
reported in Ref.\onlinecite{Anis13}. Magnetotransport measurements were
performed at different orientations of the magnetic field in the field range
up to 9 T using a superconducting solenoid. The field orientation was
defined by the angle $\theta $ between the field direction and the normal $b$%
-axis to the highly conducting $(a,c)$ plane. We used a homemade rotator
with an angular accuracy better than 0.1$^{\circ }$, having previously
allowed to demonstrate the two dimensionality behavior of BSCCO high $T_c$
superconductors \cite{Labdi97}. A great care was made to get rid off any
backlash in the rotation.

\begin{figure}[t]
\includegraphics[width=8.5cm]{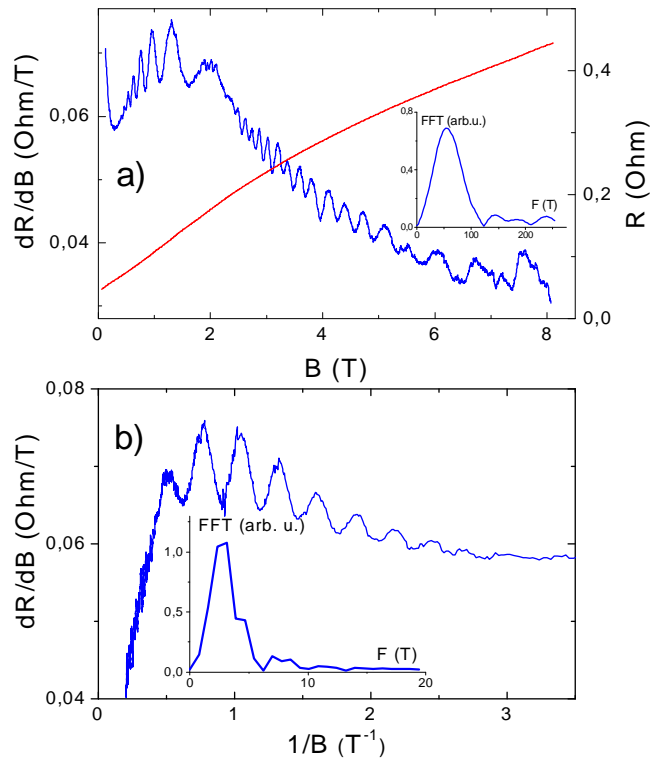}
\caption{(color online) a) magnetoresistance $R(B)$ (red curve) and $%
dR(B)/dB $ (blue curve) dependencies at 4.2 K of GdTe$_3$ demonstrating
rapid Shubnikov-de-Haas-type oscillations which appear at $B>2$ T. Inset
shows the Fourier transform of Shubnikov-de-Haas oscillations. b) variation
of $dR(B)/dB$ as a function of the inverse magnetic field, $B^{-1}$, in the
low field range $B<2$ T demonstrating slow oscillations (SlO). Inset shows
the corresponding Fourier transform of SlO.}
\label{F1}
\end{figure}

\begin{figure}[t]
\includegraphics[width=8.5cm]{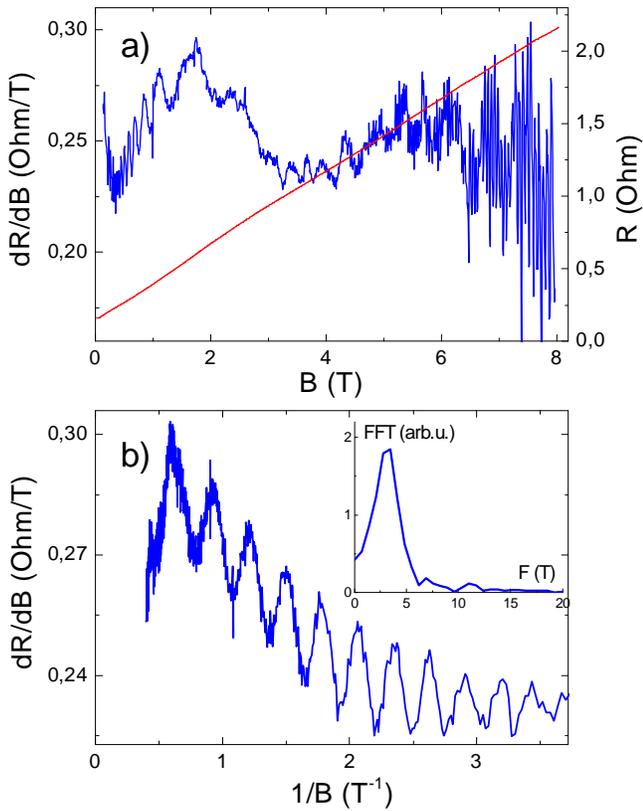}
\caption{(color online) the same as in Fig.\protect\ref{F1} for TbTe$_3$.}
\label{F2}
\end{figure}

The magnetoresistance $R(B)$ and its derivative $dR(B)/dB$ as a function of
the magnetic field up to $B=8.2$ T applied along the $b$-axis and with the
current applied in the $(a,c)$ plane at $T=4.2$ K are drawn in Fig.\ref{F1}a
for GdTe$_3$ and in Fig.\ref{F2}a for TbTe$_3$. For both compounds,
oscillations with a very weak amplitude are detectable. At $B>2$ T
pronounced Shubnikov-de-Haas (SdH) oscillations with a frequency $%
F\approx55-58$ T are observed in $dR/dB$ as seen in the inset of Fig \ref{F1}%
a for GdTe$_3$. At high field ($B\gtrsim7$ T) new oscillations with high
frequency ($F\approx 0.7-0.8$ kT) appear in TbTe$_3$, indicating the
existence of several types of pockets on the partially gapped Fermi surface
(FS). De Haas-van Alphen oscillations were previously observed\cite{LaTe08}
from a.c. susceptibility and torque measurements in LaTe$_3$ with three
distinct frequencies $\alpha\sim 50$ T, $\beta\approx 520$ T and $\gamma\sim
1600$ T. The $\beta$ frequency was attributed to small FS pockets around the
$X$ point in the Brillouin zone, unaffected by the CDW, while the $\alpha $
frequency was assigned to a portion of the reconstructed FS. We can
attribute the observed frequency $F\approx 56$ T of SdH oscillations above 2
T in GdTe$_3$ and TbTe$_3$ similarly to the $\alpha$ frequency in LaTe$_3$%
\cite{LaTe08}.

However, the more striking result, shown in Figs. \ref{F1}a and \ref{F2}a,
is that, in addition to the rapid SdH oscillations, at low magnetic field ($%
B<2$ T) the magnetoresistance exhibits prominent slow oscillations (SlO)
with a very low frequency $F_{slow}\lesssim 4$ T. In Figs. \ref{F1}b and \ref%
{F2}b, we have plotted the derivative $dR(B)/dB$ as a function of inverse
magnetic field with its Fourier transform (FFT) in the insets. The FFT of
slow oscillations and of usual quantum oscillations was done in different
magnetic field ranges. Therefore, two peaks of FFT at about $3.5$ and $56$ T
in the spectrum appear only on different plots in Fig. \ref{F1}a and Fig. %
\ref{F1}b. Below we focus specifically on these slow oscillations.

\begin{figure}[t]
\includegraphics[width=8cm]{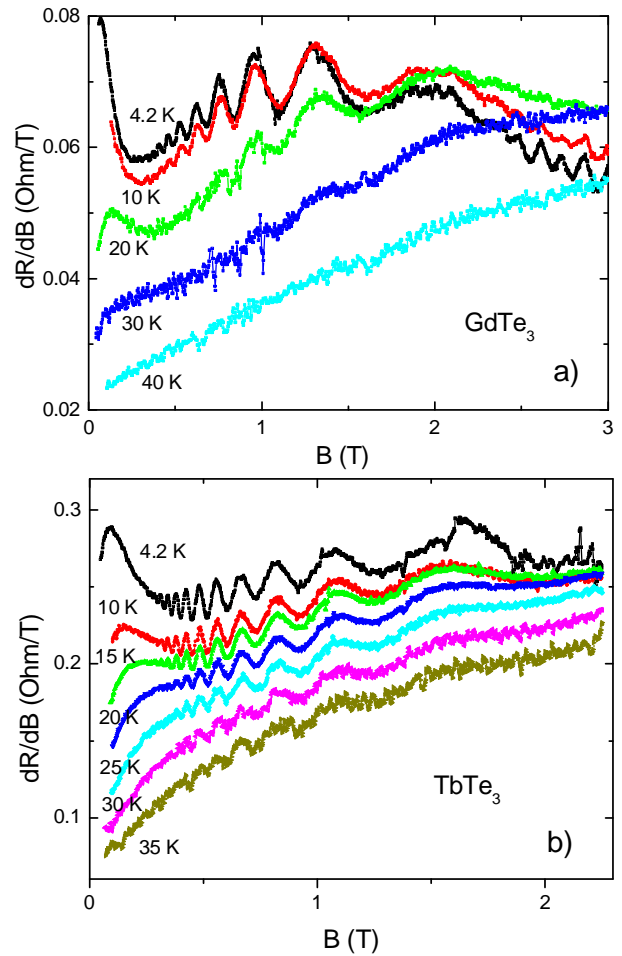}
\caption{(color online) Temperature evolution of slow oscillations in
magnetoresistance for GdTe$_{3}$ (a) and TbTe$_{3}$ (b).}
\label{F3}
\end{figure}

In contrast to the usual SdH oscillations, the amplitude of which decreases
rapidly as temperature increases, the SlO of MR are observable up to $%
T\backsimeq 40$ K, as can be seen from Fig. \ref{F3} where we show the
temperature evolution of SlO for GdTe$_{3}$ and TbTe$_{3}$. If one extract
the electron effective mass from such weak temperature dependence of SlO
amplitude, one obtains $m^{\ast }\approx 0.004m_{e}$, which is unreasonably
small. This suggests that the observed SlO originate not from small FS
pockets, but from the FS warping due to $t_{z}$, similarly to the SlO of
interlayer MR in the organic superconductor $\beta $-(BEDT-TTF)$_{2}$IBr$%
_{2} $ \cite{SO}, or due to the bilayer splitting $t_{b}$. If so, the
observed SlO give an excellent opportunity to measure the values of $t_{b}$
or $t_{z}$ and $k_{F}$ at low temperature in rare-earth tritellurides TbTe$%
_{3}$ and GdTe$_{3}$. To discriminate between the two possible origins of
SlO, in the next section we consider them in more detail.

\section{Theoretical description}

\subsection{Slow oscillations of intralayer magnetoresistance due to
interlayer dispersion}

According to Eq. (90.5) of Ref. \onlinecite{LL10}, the intralayer
conductivity at finite temperature is given by
\begin{equation}
\sigma _{yy}=e^{2}\int d\varepsilon \,\left[ -n_{F}^{\prime }(\varepsilon )%
\right] \,g\left( \varepsilon \right) D_{y}\left( \varepsilon \right),
\label{sigmayy}
\end{equation}%
where the derivative of the Fermi distribution function $n_{F}^{\prime
}(\varepsilon )=-1/\{4T\cosh ^{2}\left[ (\varepsilon -\mu )/2T\right] \}$, $%
g\left( \varepsilon \right) $ is the DoS and $D_{y}\left( \varepsilon
\right) $ is the diffusion coefficient of electrons along y-axis. Below one
only needs the first terms in the harmonic expansion for the oscillating
DoS, which in Q2D metals at finite $t_{z}\sim \hbar \omega _{c}$ are given
by \cite{Champel2001,ChampelMineev,Shub}%
\begin{equation}
g\left( \varepsilon \right) \approx g_{0}\left[ 1-2\cos \left( \frac{2\pi
\varepsilon }{\hbar \omega _{c}}\right) J_{0}\left( \frac{4\pi t_{z}}{\hbar
\omega _{c}}\right) R_{D}\right] ,  \label{ge1}
\end{equation}
where $g_{0}=m^{\ast }/\pi \hbar ^{2}d$ is the DoS at the Fermi level in the
absence of magnetic field per two spin components\cite{CommentDoS}, $%
J_{0}\left( x\right) $ is the Bessel's function, the Dingle factor\cite%
{Dingle,Bychkov} $R_{D}\approx \exp \left[ -\pi k/\omega _{c}\tau _{0}\right]
$, $\tau _{0}$ is the electron mean free time without magnetic field.

To calculate the diffusion coefficient\cite{CommentDiff} $D_{y}\left(
\varepsilon \right) $, we consider only short-range impurities, described by
a $\delta $-function potential: $V_{i}\left( r\right) =U\delta ^{3}\left(
r-r_{i}\right) $. The matrix element of impurity scattering is given by $%
T_{mm^{\prime }}=\Psi _{m^{\prime }}^{\ast }\left( r_{i}\right) U\Psi
_{m}\left( r_{i}\right) $, where $\Psi _{m}\left( r\right) $ is the electron
wave function in the state $m$. During each scattering, the typical change $%
\Delta y=\Delta P_{x}c/eB_{z}$ of the mean electron coordinate $y_{0}$
perpendicular to $\boldsymbol{B}$ is of the order of Larmor radius $%
R_{L}=p_{F}c/eB_{z}$.\cite{CommentDecay,Fogler1997,Fogler1998} The diffusion
coefficient is approximately given by
\begin{equation}
D_{y}\left( \varepsilon \right) \approx \left\langle \left( \Delta y\right)
^{2}\right\rangle /2\tau \left( \varepsilon \right) ,  \label{D1}
\end{equation}%
where $\tau \left( \varepsilon \right) $ is the energy-dependent electron
mean scattering time by impurities, and the angular brackets in Eq. (\ref{D1}%
) mean averaging over impurity scattering events. In the Born approximation,
the mean scattering rate $1/\tau \left( \varepsilon \right) =2\pi
n_{i}U^{2}g\left( \varepsilon \right) $, where $n_{i}$ is the impurity
concentration. This scattering rate has MQO, proportional to those of the
DoS in Eq. (\ref{ge1}). The MQO of $\left\langle \left( \Delta y\right)
^{2}\right\rangle \approx R_{L}^{2} $ are, usually, weaker and in 3D metals
they are neglected \cite{LL10}. Then $D_{y}\left( \varepsilon \right)
\approx R_{L}^{2}/2\tau \left( \varepsilon \right) \propto g\left(
\varepsilon \right) $. However, in Q2D metals, when $t_{z}\sim \hbar \omega
_{c}$, the MQO of $\left\langle \left( \Delta y\right) ^{2}\right\rangle $
can be of the same order as the MQO of the DoS, and at $R_{D}\ll 1$
\begin{equation}
D_{y}\left( \varepsilon \right) \approx D_{0}\left[ 1-2\alpha \cos \left(
\frac{2\pi \varepsilon }{\hbar \omega _{c}}\right) J_{0}\left( \frac{4\pi
t_{z}}{\hbar \omega _{c}}\right) R_{D}\right] ,  \label{D3}
\end{equation}%
where $D_{0}\approx R_{L}^{2}/2\tau _{0}$, and the number $\alpha \sim 1$.
Combining Eqs. (\ref{sigmayy}),(\ref{ge1}) and (\ref{D3}) after the
integration over $\varepsilon $ we obtain
\begin{gather}
\frac{\sigma _{yy}(B)}{e^{2}g_{0}D_{0}}\approx 1+2\alpha J_{0}^{2}\left(4\pi
t_{z}/\hbar \omega _{c}\right) R_{D}^{2}-  \label{SO} \\
-2\left( \alpha +1\right) \cos \left( \frac{2\pi \mu }{\hbar \omega _{c}}%
\right) J_{0}\left( \frac{4\pi t_{z}}{\hbar \omega _{c}}\right) R_{D}R_{T},
\notag
\end{gather}
where the temperature damping factor of the MQO is
\begin{equation}
R_{T}=\left( 2\pi ^{2}k_{B}T/\hbar \omega _{c}\right) /\sinh \left( 2\pi
^{2}k_{B}T/\hbar \omega _{c}\right) .  \label{RT}
\end{equation}%
The temperature damping factor (\ref{RT}) in the second MQO term in Eq. (\ref%
{SO}) arises from the integration over energy $\varepsilon $ of the rapidly
oscillating function $\propto \cos ( 2\pi \varepsilon / \hbar \omega _{c})$
with the derivative of Fermi distribution function $n_{F}^{\prime }$
according to Eq. (\ref{sigmayy}). The SlO term arises from the $\varepsilon $%
-independent product $J_{0}^{2}( 4\pi t_{z}/\hbar \omega _{c}) $, and its
integration over $\varepsilon $ in Eq. (\ref{sigmayy}) does not produce the
temperature damping factor (\ref{RT}). Hence, the SlO, described by the
first line of Eq. (\ref{SO}), are not damped by temperature within our
model, similarly to Refs. \cite{SO,Shub}.

Approximately, one can use the asymptotic expansion of the Bessel function
in Eq. (\ref{SO}) for large values of the argument: $J_{0}(x)\approx \sqrt{%
2/\pi x}\cos \left(x-\pi/4\right)\,,\,x\gg 1$. Then, after introducing the
frequency of the SlO, $F_{slow}= 4t_{z}B/\hbar \omega
_{c}=4t_{z}m^{\ast}c/e\hbar $, the first line in Eq. (\ref{SO}) simplifies
to
\begin{equation}
\frac{\sigma ^{slow}_{yy}(B)}{e^{2}g_{0}D_{0}}\approx 1+\frac{\alpha \hbar
\omega _{c}}{2\pi ^{2}t_{z}}\sin \left( \frac{2\pi F_{slow}}{B}\right)
R_{D}^{2}.  \label{SOa}
\end{equation}

In tilted magnetic field at constant $\left\vert \boldsymbol{B}\right\vert $%
, $\omega _{c}\propto \cos \theta $ and the angular dependence of interlayer
transfer integral is\cite{Kur}
\begin{equation}
t_{z}\left( \theta \right) =t_{z}\left( 0\right) J_{0}\left( k_{F}d\tan
\theta \right) ,  \label{tzTheta}
\end{equation}%
where $d$ is the interlayer distance. Then the frequency of the SlO must
depend on the tilt angle $\theta $ as:
\begin{equation}
F_{slow}\left( \theta \right) /F_{slow}\left( 0\right) =J_{0}\left(
k_{F}d\tan \theta \right) /\cos \left( \theta \right) .  \label{Fslow}
\end{equation}%
Eqs. (\ref{tzTheta}) and (\ref{Fslow}) assume a single value of the in-plane
Fermi momentum $k_{F}$. If there are several different FS pockets, the slow
oscillations are given by a sum of the contributions from each pocket. Then
the simple angular dependence in Eq. (\ref{Fslow}) is smeared out, and the
deep minima of the SlO frequency $F_{slow}\left( \theta \right) $ at the
Yamaji angles, observed in Ref. \cite{SO}, become weaker or even disappear,
being only seen as a splitting or just as a broadening of the Fourier
transform peak at certain angles $\theta $. The similar smearing of the
simple dependence in Eq. (\ref{Fslow}) occurs when the FS pockets are
elongated and oriented differently. On the other hand, if the product of
interlayer transfer integral $t_{z}$ and cyclotron mass $m^{\ast }$ is the
same for all FS pockets, all FS pockets contribute to SlO with the same
frequency, which additionally enhances the SlO amplitude as compared to MQO
amplitudes. More probable is the case when the SlO frequencies from
different FS pockets are close but do not coincide exactly, which enhances
but broadens the SlO peak in the Fourier transform of magnetoresistance.

\subsection{Slow oscillations due to bilayer splitting}

Another possible origin of the slow oscillations comes from the entanglement
of two close frequencies due to the bilayer splitting. The elementary
crystal cell of RTe$_{3}$ in the interlayer z-direction has two conducting
Te bilayers separated by insulating RTe slabs (see Fig. 1 in Refs. %
\onlinecite{LaTe08} and \onlinecite{SchmittNJP2011}). The interlayer
distances are well know for the close compound NdTe$_{3}$.\cite{N1966} In
NdTe$_{3}$ the Te layers within one bilayer are separated by a distance of
only $d^{\star }\approx 3.64\mathring{A}$, and the bilayers are separated by
$h\approx 9.26\mathring{A}$.\cite{N1966} As a result the lattice constant $%
c^{\star }=2(h+d^{\star })\approx 25.8$\AA\ in the interlayer z-direction in
RTe$_{3}$ is very large. We take these values of $d^{\star }$,$c^{\star }$
and $h$ for our study of TbTe$_{3}$ and GdTe$_{3}$.

Assume that the coupling $t_{z}$ between the bilayers, leading to the
interlayer $k_{z}$ energy dispersion, is negligibly weak, and consider only
one bilayer. The interlayer hopping $t_{b}$ between adjacent layers within
one bilayer leads to the so-called bonding and anti-bonding energy states,
respectively corresponding to the even and odd electron wave functions in
the z-direction. The energy of bonding (even) state is lower than the energy
of antibonding (odd) state by the value $\Delta \epsilon \approx 2t_{b}$.
This bilayer splitting is very common also in high-temperature cuprate
BISCCO and YBCO superconductors, where it has been extensively studied.\cite%
{FengBilayer2001,GarciaNJP2010,HarrisonSciRep2015} The slow oscillations due
to bilayer splitting in combination with $k_{z}$ dispersion allowed to
explain the three close slow frequencies of MQO observed in YBCO.\cite%
{SlowYBCO} For us it is important only that the in-plane Fermi energy of
bonding states is higher than the Fermi energy of antibonding states by this
energy splitting $\Delta \epsilon \approx 2t_{b}$. This results in the
corresponding splitting of the basic frequency $F_{0}$ of MQO: $%
F_{0}\rightarrow $ $F_{0}\pm \Delta F$. Then the DoS is given by a sum of
the bonding and antibonding states, and instead of Eq. (\ref{ge1}) for the
DoS we then obtain%
\begin{equation}
\frac{g\left( \varepsilon \right) }{g_{0}}\approx 1-R_{D}\cos \left( 2\pi
\frac{\varepsilon +t_{b}}{\hbar \omega _{c}}\right) -R_{D}\cos \left( 2\pi
\frac{\varepsilon -t_{b}}{\hbar \omega _{c}}\right) ,  \label{gb1}
\end{equation}%
where $g_{0}$ is DoS for two layers (one bilayer). Similarly, instead of Eq.
(\ref{D3}) for the diffusion coefficient we obtain
\begin{equation}
\frac{D_{y}\left( \varepsilon \right) }{D_{0}}\approx 1-\alpha R_{D}\left[
\cos \left( 2\pi \frac{\varepsilon +t_{b}}{\hbar \omega _{c}}\right) +\cos
\left( 2\pi \frac{\varepsilon -t_{b}}{\hbar \omega _{c}}\right) \right] .
\label{Db1}
\end{equation}%
Instead of Eq. (\ref{SO}) for the intralayer conductivity from Eq. (\ref%
{sigmayy}) one then obtains
\begin{gather}
\frac{\sigma _{yy}(B)}{e^{2}g_{0}D_{0}}\approx 1+\alpha \cos \left( \frac{%
4\pi t_{b}}{\hbar \omega _{c}}\right) R_{D}^{2}-  \label{SOb} \\
-\left( \alpha +1\right) \cos \left( \frac{2\pi \mu }{\hbar \omega _{c}}%
\right) \cos \left( \frac{2\pi t_{b}}{\hbar \omega _{c}}\right) R_{D}R_{T}.
\notag
\end{gather}%
The SlO, described by the first line of Eq. (\ref{SOb}), are not damped by
temperature within our model again, similarly to Refs. \cite{SO,Shub} and
Eq. (\ref{SO}). However, there are several important differences of SlO
arising from FS warping and from bilayer splitting. In contrast to the case
of FS warping due to $k_{z}$ dispersion, the frequency of the SlO in the
case of bilayer splitting is given by $F_{slow}=2t_{b}B/\hbar \omega _{c}$
not only at $4\pi t_{b}\gg \hbar \omega _{c}$ but at any ratio $t_{b}/\hbar
\omega _{c}$. Also, contrary to Eq. (\ref{SOa}), the SlO amplitude in Eq. (%
\ref{SOb}) for the case of bilayer splitting does not have the small factor $%
J_{0}^{2}\left( 4\pi t_{z}/\hbar \omega _{c}\right) \sim \hbar \omega
_{c}/4\pi ^{2}t_{z}$. The phase of slow oscillations due to bilayer
splitting $t_{b}$ in Eq. (\ref{SOb}) is shifted by $\pi /2$ as compared to
the phase in Eq. (\ref{SOa}) of SlO due to $k_{z}$ dispersion.

Probably, most evident difference between the SlO due to bilayer splitting
and due to $k_{z}$ dispersion is in the angular dependence of the SlO
frequency. This SlO frequency $F_{slow}\left( \theta \right) $ does not
necessarily obey Eq. (\ref{Fslow}) but may have standard cosine dependence $%
F_{slow}\left( \theta \right) =F_{slow}\left( 0\right) /\cos \left( \theta
\right) $.\cite{Commenttb,Commenttz} Even if one assumes that Eq. (\ref%
{Fslow}) is valid also for the SlO frequency from the bilayer splitting, the
interlayer distance $d^{\star }$ in this dependence for bilayer splitting is
several times smaller than the lattice constant in interlayer z-direction.
For example, for RTe$_{3}$ compounds the lattice constant in z-direction is $%
c^{\star }=25.8$\AA , while the interlayer distance within one bilayer is
only $d^{\star }=3.64$\AA , i.e. 7 times less. Therefore, even according to
Eq. (\ref{Fslow}), the angular dependence of the frequency $F_{slow}(\theta
) $ of SlO originating from the bilayer splitting $t_{b}$ should be much
weaker than that form interbilayer coupling $t_{z}$ and should start from
much higher tilt angle $\theta $.

If there are both types of interlayer coupling, i.e. the transfer integral $%
t_{b}=t_{b}(\boldsymbol{k}_{\parallel })$ between adjacent layers separated
by distance $d^{\star }$ within one bilayer and the hopping $t_{z}=t_{z}(%
\boldsymbol{k}_{\parallel })$ between adjacent equivalent bilayers,
separated by distance $h$, where $\boldsymbol{k}_{\parallel }$ is the
intralayer momentum, the resulting electron energy spectrum is given by
(see, e.g., Eq. (6) of Ref. \onlinecite{GarciaNJP2010})
\begin{equation}
\epsilon _{\pm }\left( k_{z},\boldsymbol{k}_{\parallel }\right) =\epsilon
_{\parallel }\left( \boldsymbol{k}_{\parallel }\right) \pm \sqrt{%
t_{z}^{2}+t_{b}^{2}+2t_{z}t_{b}\cos \left[ k_{z}\left( h+d^{\star }\right) %
\right] }.  \label{EpsBilayerKz}
\end{equation}%
For $t_{z}\ll t_{b}$ this equation just gives the double bilayer splitting
to bonding and antibonding states. Note, that the derivation of Eq. (\ref%
{EpsBilayerKz}) assumes\cite{GarciaNJP2010} that all bilayers are
equivalent, i.e. that the lattice constant in z-direction $c^{\star
}=h+d^{\star }$. If the bilayers are nonequivalent, as in the case of RTe$%
_{3}$ compounds where $c^{\star }=2(h+d^{\star })$, Eq. (\ref{EpsBilayerKz})
needs further modification, which is the subject of separate publication.
However, we should notice that if the observed slow oscillations in RTe$_{3}$
are due to the coupling $t_{z}$ between bilayers, in the angular dependence
in Eqs. (\ref{tzTheta}) and (\ref{Fslow}) the distance $h+d^{\star
}=c^{\star }/2$ between adjacent bilayers rather than the total lattice
constant $c^{\star }$ enters as the interlayer distance $d$.

\section{Discussion}

\begin{figure}[t]
\includegraphics[width=8cm]{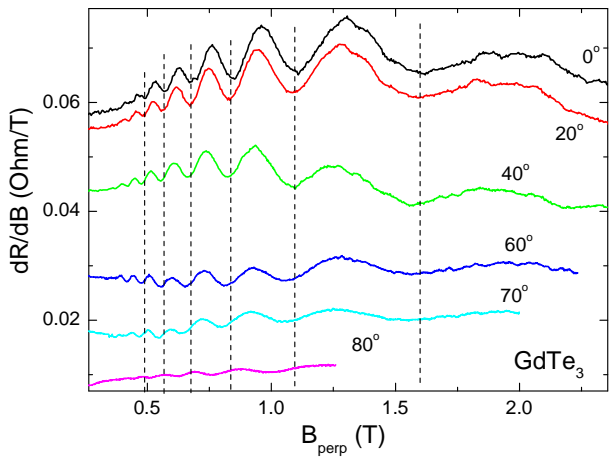}
\caption{(color online) Slow oscillations observed in GdTe$_{3}$ at $T=4.2$
K for different tilt angles $\protect\theta $ between the magnetic field $%
\boldsymbol{B}$ and the normal to the conducting layers. $B_{\perp }=B\cos (%
\protect\theta )$.}
\label{F4}
\end{figure}

To clarify the origin of the observed SlO, we have experimentally studied
the angular dependence of the SlO frequency. The evolution of the SlO in GdTe%
$_{3}$ with the change of the tilt angle $\theta $ of magnetic field at $%
T=4.2$ K is shown in Fig. \ref{F4}, where the derivative $dR/dB$ is plotted
as a function of the perpendicular-to-layers component of the magnetic field
$B_{\perp }=B\cos (\theta )$. Note that the magnetic field rotation in the ($%
b$-$c$) and ($b$-$a$) planes demonstrated the same results for TbTe$_{3}$.

\begin{figure}[t]
\includegraphics[width=8.5cm]{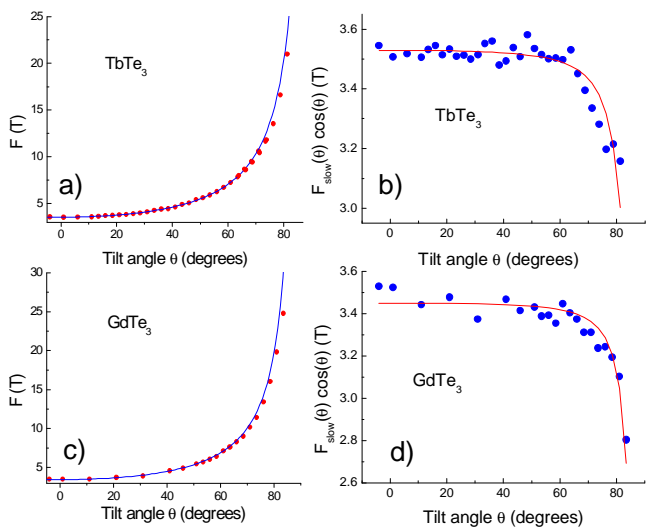}
\caption{(color online) (a) and (c): The frequency of the slow oscillations
(SlO) $F_{slow}$ as a function of tilt angle $\protect\theta $ at $T=4.2$K
for TbTe$_{3}$ and GdTe$_{3}$. Solid curves show the function $F(\protect%
\theta)=F(0)/\cos (\protect\theta )$. (b) and (d): The angular dependence of
the SlO frequency $F_{slow}(\protect\theta )$ in the intralayer
magnetoresistance in TbTe$_{3}$ and GdTe$_3$ correspondingly, multiplied by $%
\cos (\protect\theta )$. The experimental data are shown by blue filled
circles, and the theoretical prediction according to Eq. (\protect\ref{Fslow}%
) with $k_{F}d=0.12$ is shown by solid red lines.}
\label{F5}
\end{figure}

In Fig. \ref{F5} we show the $\theta $-dependence of the SlO frequency $%
F_{slow}$ at $T=4.2$ K for TbTe$_{3}$ (a) and for GdTe$_{3}$ (c). The solid
curves give the cosine dependence $F(\theta )=F(0)/\cos (\theta )$ typical
for MQO. According to Eq. (\ref{Fslow}), $F_{slow}\left( \theta \right) $
differs from this standard cosine dependence, especially at high tilt angle.
In Fig. \ref{F5} (b) we plot the angular dependence of the product $%
F_{slow}(\theta )\cos (\theta )$ in TbTe$_{3}$. If the origin of the SlO was
due to small FS pockets, the product $F_{slow}(\theta )\cos (\theta )$ would
be independent of the tilt angle $\theta $. The experimental data, shown by
blue filled circles, clearly indicate the deviation from the horizontal
line. These experimental data can be reasonably fitted by Eq. (\ref{Fslow})
with $k_{F}d=0.11$, shown by solid red lines in Figs. \ref{F5} (b,d). This
supports our assertion that the observed slow oscillations originate not
from small FS pockets as usual SdH oscillations, but from the entanglement
of close frequencies due to a finite interlayer hopping $t_{z}$ or $t_{b}$.
Another argument in favor of this origin of the observed SlO is the very
weak temperature dependence of their amplitude. To our knowledge, the data
obtained are the first observation of such SlO in the intralayer
magnetotransport.

The third argument, supporting the proposed origin of SlO as due to the
interlayer hopping rather than due to very small ungapped FS pockets, is
that the frequency of the observed SlO is independent of temperature.
Indeed, if the observed SlO originated from very small ungapped FS pockets,
their frequency would strongly depend on temperature on the scale of the CDW
transition temperature, because the size of the ungapped FS pockets depends
on the temperature-dependent CDW energy gap. For TbTe$_{3}$ the second CDW
transition temperature is\cite{Banerjee} $T_{c2}$=41K, but we do not observe
any change in the frequency of SlO up to 35K (see Fig. \ref{F2}), which is
inconsistent with the small FS-pocket origin of SlO. On contrary, the
interlayer transfer integrals $t_{z}$ or $t_{b}$ are not sensitive to the
in-plane electronic phase transitions and to the in-plane Fermi-surface
reconstruction. The interlayer transfer integrals $t_{z}$ and $t_{b}$ are
determined mainly by the strong ($\sim $1eV) crystalline potential in the
interlayer direction, which is not affected by the CDW or other in-plane
electronic orderings.

According to Eq. (\ref{Fslow}), the angular dependence of the frequency $%
F_{slow}\left( \theta \right) $ of SlO allows to estimate the value of the
Fermi momentum of the open FS pockets.\cite{SO} Fitting the experimental
data of $F_{slow}\left( \theta \right) $ shown in Fig. \ref{F5} to Eq. (\ref%
{Fslow}) gives $k_{F}d\approx 0.11$ for GdTe$_{3}$ and $k_{F}d\approx 0.12$
for TbTe$_{3}$. As we showed before, there are two possible origins of the
observed SlO in RTe$_{3}$: the bilayer splitting $t_{b}$ and the
inter-bilayer coupling $t_{z}$. The first double splits the Fermi energy,
while the latter leads to the $k_{z}$ energy dispersion and to the FS
warping. Correspondingly, there are two interlayer distances: $d^{\star
}\approx 3.64$\AA\ and $c^{\star }/2=h+d^{\star }\approx 12.9$\AA . With $%
d=d^{\star }=3.64$\AA\ we obtain $k_{F}\approx 3.3\cdot 10^{6}cm^{-1}$, and
with $d=c^{\star }/2=h+d^{\star }\approx 12.9$\AA\ we obtain $k_{F}\approx
9.3\cdot 10^{5}cm^{-1}$. If one assumes that these small FS pockets are not
elongated\cite{CommentElongatedFS} but almost circular, the corresponding FS
cross section areas are $S_{ext}\approx \pi k_{F}^{2}$. For the obtained
value $k_{F}\approx 3.3\cdot 10^{6}cm^{-1}$ for bilayer splitting ($%
d=d^{\star }$) this gives the MQO frequency $F_{0}=S_{ext}\hbar c/2\pi
e\approx 36T$, a value close to the frequency 55-58 T of oscillations we
have measured (inset of Fig. \ref{F1}a). The difference between the
estimated 36 T and the experimental value 55-58 T can be accounted by
considering the elongation or another non-circular shape of the FS pockets.%
\cite{CommentElongatedFS} Thus, the scenario of the bilayer-splitting origin
of SlO looks self-consistent. On the other hand, for the FS warping origin
of SlO, taking $d=h+d^{\star }$ and $k_{F}\approx 9.3\cdot 10^{5}cm^{-1}$
gives only $F_{0}\approx 3T$. Such a small fundamental frequency of MQO was
not measured. Thus the observed angular dependence of SlO frequency suggests
that the observed SlO originate from bilayer splitting $t_{b}$ rather than
from FS warping due to $t_{z}$.

\begin{figure}[t]
\includegraphics[width=8cm]{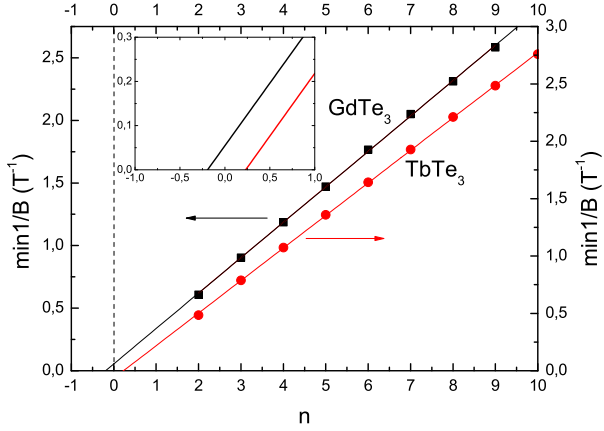}
\caption{(color online) The measured positions $1/B_{min}$ of the minima in
the derivative $dR/dB$ for GdTe$_{3}$ (black squares) and TbTe$_{3}$ (red
circles) at $T=4.2$ K as function of the number $n$ of these minima. The
experimental data are taken from Figs. \protect\ref{F1} and \protect\ref{F2}%
. The solid lines are the best linear fits. Insert figure shows the region
around $n=0$ in a larger scale to emphasize that the fitting lines intersect
abscissa axis at $\pm 1/4$.}
\label{F6}
\end{figure}

To further clarify the origin of the observed SlO, we now analyze their
phase, which depends on the origin of SlO. In the first scenario, when the
SlO originate from the FS warping and interbilayer coupling $t_{z}$, the SlO
are described by Eq. (\ref{SOa}). At small magnetic field $B<2\,$T, when SlO
are observed, the Hall conductivity $\sigma _{xy}\ll \sigma _{yy}$, and the
diagonal magnetoresistance $R_{yy}=\sigma _{xx}/\left( \sigma _{xx}\sigma
_{yy}-\sigma _{xy}^{2}\right) \approx 1/\sigma _{yy}$. Then from Eq. (\ref%
{SOa}) one obtains that the derivative $dR/dB$, shown in Figs. \ref{F1}b and %
\ref{F2}b, is approximately given by
\begin{equation}
\frac{dR_{yy}^{slow}(B)}{dB}\propto 1+\frac{\alpha \hbar \omega _{c}F_{slow}%
}{\pi t_{z}B^{2}}\cos \left( \frac{2\pi F_{slow}}{B}\right) R_{D}^{2},
\label{dRdBW}
\end{equation}%
and the position $B_{\min ,W}\left( n\right) $ of the $n$-th minimum of SlO
of $dR(B)/dB$ for the warping scenario of SlO is given by%
\begin{equation}
F_{slow}/B_{\min ,W}\left( n\right) =n-1/2.  \label{BminW}
\end{equation}

In the second scenario, when SlO originate from the bilayer splitting $t_{b}$%
, one should apply Eq. (\ref{SOb}) instead of Eq. (\ref{SOa}), which gives $%
R_{yy}(B)\propto 1-\alpha \cos \left( 2\pi F_{slow}/B\right) R_{D}^{2}$ and
\begin{equation}
\frac{dR_{yy}(B)}{dB}\propto 1-\alpha \frac{2\pi F_{slow}}{B^{2}}\sin \left(
\frac{2\pi F_{slow}}{B}\right) R_{D}^{2}.  \label{dRdBB}
\end{equation}%
The position $B_{\min ,b}\left( n\right) $ of the $n$-th minimum of $%
dR(B)/dB $ in Eq. (\ref{dRdBB}) is given by
\begin{equation}
F_{slow}/B_{\min ,b}\left( n\right) =n+\mathrm{sign}\left( \alpha \right) /4.
\label{BminB}
\end{equation}

The experimental data on the phase of SlO are shown in Fig. \ref{F6} and can
be well fitted by Eq. (\ref{BminB}), corresponding to the bilayer-splitting
origin of SlO. On contrary, these data cannot be fitted by Eq. (\ref{BminW}%
), corresponding to the FS-warping scenario of SlO, originating from the
interbilayer coupling $t_{z}$. However, it is not clear why the phase offset
$1/4$ in Fig. \ref{F6} for GdTe$_{3}$ and TbTe$_{3}$ has different sign,
formally corresponding to the different sign of the coefficient $\alpha $.
This difference may, in principle, appear if the reconstructed FS or the
parameter $\omega _{c}\tau $ is considerably different for these two
compounds. Therefore, a more rigorous calculation of $\alpha $ in terms of
the initial parameters $\omega _{c}\tau $ and $t_{b}/\hbar \omega _{c}$ and
detailed experimental data on MQO in these two compounds are needed for
understanding this difference.

The observed angular dependence of the frequency $F_{slow}\left( \theta
\right) $ and the phase of SlO are both in favour of the bilayer-splitting
origin of SlO. There is a third argument, supporting this conjecture. If the
observed SlO with frequency $F\approx 4T$ were due to FS warping and
inter-bilayer hopping $t_{z}$, one would expect to observe another SlO with
larger frequency, corresponding to the bilayer splitting and the transfer
integral $t_{b}>t_{z}$. According to Eqs. (\ref{SOa}) and (\ref{SOb}), the
SlO from bilayer splitting should have larger amplitude than SlO from FS
warping because of the extra factor $\hbar \omega _{c}/2\pi ^{2}t_{z}$ in
Eq. (\ref{SOa}) as compared to Eq. (\ref{SOb}). Thus, the second SlO would
have even larger amplitude than the observed SlO. However, on experiment
there is no any signature of the second SlO, which supports our assertion
that the observed SlO originate from the bilayer splitting $t_{b}$ rather
than from FS warping $t_{z}$. To our knowledge, the reported results are the
first experimental and theoretical study of the slow oscillations of MR
originating from the bilayer splitting. However, this phenomenon is expected
to be rather general and should be observable in many other bilayered
materials.

The SlO of intra- and interlayer electron transport, studied above and in
Refs. \cite{SO} and \cite{Shub}, are qualitatively similar and have only
some minor quantitative differences in amplitude and phase (compare Eq. (\ref%
{SOa}) above with Eq. (4) of Ref. \cite{SO}). On the other hand, the SlO
originating from FS warping and from bilayer splitting have qualitative
differences, e.g. in the angular dependence of SlO frequency.

The frequency of the SlO at $\theta =0$ can be used to estimate the value $%
t_{b}$ of the interlayer transfer integral. According to Eq. (\ref{SOa}),
with the effective electron mass $m^{\ast }\approx 0.1m_{e}$ determined from
the temperature dependence of the amplitude of SdH oscillations \cite{JLTP15}%
, and $F_{slow}\approx 3.5$T (see Fig. \ref{F5} (b,d)), one obtains $%
t_{b}\approx 1meV$. These small values of the interlayer transfer integral $%
t_{b}$ in comparison to much larger intralayer transfer integrals $%
t_{\parallel }\approx 2$ eV along the chains and $t_{\perp }\approx 0.37$ eV
perpendicular to the chains in the $(a,c)$ plane, as obtained by the band
structure calculations \cite{Brouet08}, illustrate the quasi-2D character of
these rare-earth tritellurides and justify that the dispersion along the $b$%
-axis is neglected in ARPES measurements.\cite{CommentARPESLowT} The value
of interlayer transfer integral $t_{b}$ is very important for various
physical properties of strongly anisotropic compounds. The quantum
corrections to conductivity \cite{AABook,Kennett2008} rapidly decrease with
increasing of $t_{b}$, being much stronger in 2D electronic systems. The
quantum Hall effect also requires an exponentially small value of interlayer
hopping integral \cite{CommentQHE,HuckesteinRMP1995}.

The proposed technique to measure the electronic structure, namely, the
interlayer electron hopping rate and the in-plane Fermi momentum, may be
very useful to many other layered materials, including the cuprate and
Fe-base high-temperature superconductors. Probably, the quantitative theory
of slow oscillations in these materials must include the effects of strong
electronic correlations, which are missed in the present one-electron
approach \cite{CommentManyE}. However, the reported first observation and
simplified qualitative description of the slow oscillations of the in-plane
electronic magnetotransport, as well as their application to extract the
electronic-structure parameters of the studied materials, may stimulate
further application of this promising technique. The MQO observed in layered
high-Tc superconducting materials, usually, have very small amplitudes even
in the strongest available magnetic fields, which impedes their application
as a tool to study the electronic structure in these materials. The FS
reconstruction due to an electronic ordering at finite wave vector, e.g. a
density-wave or antiferromagnetic ordering, is known to additionally
suppress the MQO because of magnetic breakdown between different FS parts.
The SlO, being almost a classical type of magnetoresistance oscillations, do
not have these damping factors and can be clearer observed, which enhances
their potential use to investigate the electronic structure of various
strongly-correlated electronic systems.

To summarize, we report the first observation and qualitative theoretical
description of slow oscillations (SlO) of the intralayer magnetoresistance
in quasi-2D metallic compounds. These SlO are observed in rather weak
magnetic field $B<2T$ and at rather high temperature up to $T\approx 40K$,
contrary to the usual magnetic quantum oscillations, which are strongly
damped by temperature, especially in such weak field. The phase and the
angular dependence of the SlO frequency suggest that the observed SlO
originate from the bilayer splitting $t_{b}$ rather than from the FS warping
and inter-bilayer hopping $t_{z}$, contrary to their origin in the organic
metal in Ref. \cite{SO}. Such SlO due to bilayer splitting have not been
studied before. The SlO allow to measure the interlayer transfer integral
and the in-plane Fermi momentum $k_{F}$, which are difficult to measure by
other means. We obtained the values $t_{b}\approx 1meV$ in the rare-earth
tritelluride compounds TbTe$_{3}$ and GdTe$_{3}$. This method is useful to
many other layered conductors.

\medskip

\acknowledgments{The work was partially supported by RFBR (grants No. 14-02-01126-a
and 16-02-00522-a) and partially performed in the CNRS-RAS
Associated International Laboratory between CRTBT and IRE "Physical
properties of coherent electronic states in coherent matter".}

\medskip

A.A. Sinchenko, P. Monceau, and V.N. Zverev performed measurements. P.
Lejay, A. Hadj-Azzem, J. Balay, and O. Leynaud prepared the samples. P.D.
Grigoriev contributed to theoretical interpretation. All authors
participated in the discussion of the results.


\begin{thebibliography}{99}
\bibitem{MQORev} J. Wosnitza, \textit{Fermi Surfaces of Low-Dimensional
Organic Metals and Superconductors} (Springer-Verlag, Berlin, 1996); J.
Singleton, Rep. Prog. Phys. \textbf{63}, 1111 (2000).

\bibitem{OMRev} T.~Ishiguro, K.~Yamaji and G.~Saito, \emph{Organic
Superconductors}, 2nd Edition, Springer-Verlag, Berlin, 1998; \textit{The
Physics of Organic Superconductors and Conductors}, ed. by A. G. Lebed
(Springer Series in Materials Science, V. 110; Springer Verlag Berlin
Heidelberg 2008).

\bibitem{MarkReview2004} M.V. Kartsovnik, Chem. Rev. \textbf{104}, 5737
(2004).

\bibitem{KartPeschReview} M. V. Kartsovn\u{\i}k and V. G. Peschansky, Low
Temp. Phys. \textbf{31}, 185 (2005) [Fiz. Nizk. Temp. \textbf{31}, 249
(2005)].

\bibitem{HusseyNature2003} N. E. Hussey, M. Abdel-Jawad, A. Carrington, A.
P. Mackenzie and L. Balicas, Nature \textbf{425}, 814 (2003).

\bibitem{AbdelNature2006} M. Abdel-Jawad, M. P. Kennett, L. Balicas, A.
Carrington, A. P. Mackenzie, R. H. McKenzie \& N. E. Hussey, Nature Phys.
\textbf{2}, 821 (2006).

\bibitem{ProustNature2007} Nicolas Doiron-Leyraud, Cyril Proust, David
LeBoeuf, Julien Levallois, Jean-Baptiste Bonnemaison, Ruixing Liang, D. A.
Bonn, W. N. Hardy, Louis Taillefer, Nature \textbf{447}, 565 (2007).

\bibitem{AbdelPRL2007AMRO} M. Abdel-Jawad, J. G. Analytis, L. Balicas, A.
Carrington, J. P. H. Charmant, M. M. J. French, and N. E. Hussey , Phys.
Rev. Lett. \textbf{99}, 107002 (2007).

\bibitem{McKenzie2007} Malcolm P. Kennett and Ross H. McKenzie, Phys. Rev. B
\textbf{76}, 054515 (2007).

\bibitem{DVignolle2008} B. Vignolle, A. Carrington, R. A. Cooper, M. M. J.
French, A. P. Mackenzie, C. Jaudet, D. Vignolles, Cyril Proust \& N. E.
Hussey, Nature \textbf{455}, 952 (2008).

\bibitem{HelmNd2009} T. Helm, M.V. Kartsovnik, M. Bartkowiak, N. Bittner, M.
Lambacher, A. Erb, J. Wosnitza, and R. Gross, Phys. Rev. Lett. \textbf{103},
157002 (2009).

\bibitem{HelmNd2010} T. Helm, M.V. Kartsovnik, I. Sheikin, M. Bartkowiak, F.
Wolff-Fabris, N. Bittner, W. Biberacher, M. Lambacher, A. Erb, J. Wosnitza,
and R. Gross, Phys. Rev. Lett. \textbf{105}, 247002 (2010).

\bibitem{BaFeAs2011} Taichi Terashima, Nobuyuki Kurita, Megumi Tomita,
Kunihiro Kihou, Chul-Ho Lee, Yasuhide Tomioka, Toshimitsu Ito, Akira Iyo,
Hiroshi Eisaki, Tian Liang, Masamichi Nakajima, Shigeyuki Ishida, Shin-ichi
Uchida, Hisatomo Harima, and Shinya Uji, Phys. Rev. Lett. \textbf{107},
176402 (2011).

\bibitem{Graf2012} D. Graf, R. Stillwell, T. P. Murphy, J.-H. Park, E. C.
Palm, P. Schlottmann, R. D. McDonald, J. G. Analytis, I. R. Fisher, and S.
W. Tozer, Phys. Rev. B. \textbf{85}, 134503 (2012).

\bibitem{Kuraguchi2003} M. Kuraguchi, E. Ohmichi, T. Osada, Y. Shiraki,
Synth. Met. \textbf{133-134}, 113 (2003).

\bibitem{GraphiteIntercalatedNature2005} G. Csanyi, P. B. Littlewood, A. H.
Nevidomskyy, C. J. Pickard and B. D. Simon, Nature Physics \textbf{1}, 42
(2005).

\bibitem{Abrik} A.A. Abrikosov, \textit{Fundamentals of the theory of metals}%
, North-Holland, 1988.

\bibitem{Shoenberg} Shoenberg D. \textquotedblright Magnetic oscillations in
metals\textquotedblright , Cambridge University Press 1984.

\bibitem{Ziman} J. M. Ziman, \textit{Principles of the Theory of Solids},
Cambridge Univ. Press 1972.

\bibitem{KartsAMRO1988} M.V. Kartsovnik, P. A. Kononovich , V. N. Laukhin
and I. F. Shchegolev, JETP Lett. \textbf{48}, 541 (1988).

\bibitem{Yam} K. Yamaji, J. Phys. Soc. Jpn. \textbf{58}, 1520 (1989).

\bibitem{Yagi1990} R. Yagi, Y. Iye, T. Osada, S. Kagoshima, J. Phys. Soc.
Jpn. \textbf{59}, 3069 (1990).

\bibitem{Mark92} M. V. Kartsovnik, V. N. Laukhin, S. I. Pesotskii, I. F.
Schegolev, V. M. Yakovenko, J. Phys. I \textbf{2}, 89 (1992).

\bibitem{Bergemann} C. Bergemann, S. R. Julian, A. P. Mackenzie, S.
NishiZaki, and Y. Maeno, Phys. Rev. Lett. \textbf{84}, 2662 (2000).

\bibitem{GrigAMRO2010} P.D. Grigoriev, Phys. Rev. B \textbf{81}, 205122
(2010).

\bibitem{Coldea} A. I. Coldea, A. F. Bangura, J. Singleton, A. Ardavan, A.
Akutsu-Sato, H. Akutsu, S. S. Turner, and P. Day, Phys. Rev. B \textbf{69},
085112 (2004).

\bibitem{PesotskiiJETP95} R.B. Lyubovskii, S.I. Pesotskii, A. Gilevskii and
R.N. Lyubovskaya, JETP \textbf{80}, 946 (1995) [Zh. Eksp. Teor. Fiz. \textbf{%
107}, 1698 (1995)].

\bibitem{Zuo1999} F. Zuo, X. Su, P. Zhang, J. S. Brooks, J. Wosnitza, J. A.
Schlueter, Jack M. Williams, P. G. Nixon, R. W. Winter, and G. L. Gard,
Phys. Rev. B \textbf{60}, 6296 (1999).

\bibitem{W3} J. Hagel, J. Wosnitza, C. Pfleiderer, J. A. Schlueter, J.
Mohtasham, and G. L. Gard, Phys. Rev. B \textbf{68}, 104504 (2003).

\bibitem{W4} J.Wosnitza, Journal of Low Temperature Physics 146, 641 (2007).

\bibitem{Incoh2009} M. V. Kartsovnik, P. D. Grigoriev, W. Biberacher, and N.
D. Kushch, Phys. Rev. B \textbf{79}, 165120 (2009).

\bibitem{Kang} W. Kang, Y. J. Jo, D. Y. Noh, K. I. Son, and Ok-Hee Chung,
Phys. Rev. B \textbf{80}, 155102 (2009).

\bibitem{Wosnitza2002} J.Wosnitza, J. Hagel, J. S. Qualls, J. S. Brooks, E.
Balthes, D. Schweitzer, J. A. Schlueter, U. Geiser, J. Mohtasham, R. W.
Winter, and G. L. Gard, Phys. Rev. B 65, 180506(R) (2002).

\bibitem{SO} M.V. Kartsovnik, P.D. Grigoriev, W. Biberacher, N.D. Kushch, P.
Wyder, Phys. Rev. Lett. \textbf{89}, 126802 (2002).

\bibitem{WIPRB2012} P. D. Grigoriev, M. V. Kartsovnik, W. Biberacher, Phys.
Rev. B \textbf{86}, 165125 (2012).

\bibitem{WIPRB2011} P.D. Grigoriev, Phys. Rev. B \textbf{83}, 245129 (2011).

\bibitem{WIPRB2013} P.D. Grigoriev, Phys. Rev. B \textbf{88}, 054415 (2013).

\bibitem{GG2014} A.D. Grigoriev and P.D. Grigoriev, Low Temp. Phys. \textbf{%
40}, 367 (2014) [Fiz. Nizk. Temp. 40(4), 472 (2014)]; arXiv:1310.7109v2.

\bibitem{Shub} P.D. Grigoriev, Phys. Rev. B \textbf{67}, 144401 (2003)
[arXiv:cond-mat/0204270].

\bibitem{PhSh} P.D. Grigoriev, M.V. Kartsovnik, W. Biberacher, N.D. Kushch,
P. Wyder, Phys. Rev. B \textbf{65}, 060403(R) (2002).

\bibitem{Ru08} N. Ru, C. L. Condron, G. Y. Margulis, K. Y. Shin, J.
Laverock, S. B. Dugdale, M. F. Toney, and I. R. Fisher, Phys. Rev. B \textbf{%
77}, 035114 (2008).

\bibitem{DiMasi95} E. DiMasi, M. C. Aronson, J. F. Mansfield, B. Foran, and
S. Lee, Phys. Rev. B \textbf{52}, 14516 (1995).

\bibitem{Brouet08} V. Brouet, W. L. Yang, X. J. Zhou, Z. Hussain, R. G.
Moore, R. He, D. H. Lu, Z. X. Shen, J. Laverock, S. B. Dugdale, N. Ru, and
I. R. Fisher, Phys. Rev. B \textbf{77}, 235104 (2008).

\bibitem{SinchPRB12} A.A. Sinchenko, P. Lejay, and P. Monceau, Phys. Rev. B
\textbf{85}, 241104(R) (2012).

\bibitem{Anis13} A.A. Sinchenko, P.D. Grigoriev, P. Lejay, and P. Monceau,
Phys. Rev. Lett. \textbf{112}, 036601 (2014).

\bibitem{SSC14} A.A. Sinchenko, P. Lejay, O. Leynaud and P. Monceau, Solid
State Communications \textbf{188}, 67 (2014).

\bibitem{Iyeri2003} Y. Iyeiri, T. Okumura, C. Michioka, and K. Suzuki, Phys.
Rev. B \textbf{67}, 144417 (2003).

\bibitem{Ru2008} N. Ru, J.-H. Chu, and I. R. Fisher, Phys. Rev. B \textbf{78}%
, 012410 (2008).

\bibitem{Labdi97} S. Labdi, S.F. Kim, Z.Z. Li, S. Megtert, H. Raffy, O.
Laborde and P. Monceau, Phys. Rev. Lett. \textbf{79}, 1381 (1997).

\bibitem{LaTe08} N. Ru, R. A. Borzi, A. Rost, A. P. Mackenzie, J. Laverock,
S. B. Dugdale, and I. R. Fisher, Phys. Rev. B \textbf{78}, 045123 (2008).

\bibitem{LL10} E. M. Lifshitz and L. P. Pitaevskii, Course of Theoretical
Physics, Vol. 10: Physical Kinetics, (Nauka, Moscow, 2nd edition, 2002;
Pergamon Press, 1st edition, 1981).

\bibitem{Champel2001} V. M. Gvozdikov, Fiz. Tverd. Tela (Leningrad) 26, 2574
(1984) [Sov. Phys. Solid State 26, 1560 (1984)]; T. Champel and V. P.
Mineev, Phil. Magazine B \textbf{81}, 55 (2001).

\bibitem{ChampelMineev} T. Champel and V.P. Mineev, Phys. Rev. B \textbf{66}%
, 195111 (2002).

\bibitem{CommentDoS} Due to the renormalization of electron spectrum by a
CDW even in the mean-field approximation the DoS at the Fermi level may
change not so strongly as the FS geometry \cite{GrigorievPRB2008}, as
observed in ErTe$_3$ and HoTe$_3$.\cite{ErTe3DoS}

\bibitem{Dingle} R.B. Dingle, Proc. Roy. Soc. \textbf{A211,} 517 (1952).

\bibitem{Bychkov} Yu. A. Bychkov, Zh. Exp. Theor. Phys. \textbf{39}, 1401
(1960), [Sov. Phys. JETP \textbf{12}, 977 (1961)].

\bibitem{CommentDiff} The calculation of the diffusion coefficient $%
D_{y}\left( \varepsilon \right) $ is less trivial than of the DoS and
requires to specify the model of disorder. At $\mu \gg \hbar \omega _{c}$
the quasi-classical approximation is applicable. In an ideal crystal in a
magnetic field $\boldsymbol{B}$ the electrons move along the cyclotron
orbits with a fixed center and the Larmor radius $R_{L}=p_{F}c/eB_{z}$.
Without scattering the electron diffusion in the direction perpendicular to $%
\boldsymbol{B}$ is absent. The scattering by impurities changes the
electronic states and leads to the electron diffusion.

\bibitem{CommentDecay} For a short-range disorder the 2D electron wave
function in magnetic field decays exponentially at distance larger than the
Larmor radius.\cite{Fogler1997,Fogler1998} Therefore, for $\Delta y\gg R_{L}$
the matrix element $T_{mm^{\prime }}$ is exponentially small resulting from
the small overlap of the electron wave functions $\Psi _{m^{\prime }}^{\ast
}\left( r_{i}\right) \Psi _{m}\left( r_{i}\right) \sim \Psi _{m}^{\ast
}\left( r_{i}+\Delta y\right) \Psi _{m}\left( r_{i}\right) $.

\bibitem{Fogler1997} M. M. Fogler, A. Yu. Dobin, V. I. Perel, and B. I.
Shklovskii, Phys. Rev. B \textbf{56}, 6823 (1997).

\bibitem{Fogler1998} M. M. Fogler, A. Yu. Dobin, and B. I. Shklovskii, Phys.
Rev. B \textbf{57}, 4614 (1998).

\bibitem{Kur} Yasunari Kurihara, J. Phys. Soc. Jpn. \textbf{61}, 975 (1992).

\bibitem{JLTP15} A.A. Sinchenko et al., J. Low Temp. Phys. (2016), in press.

\bibitem{SchmittNJP2011} F. Schmitt, P. S. Kirchmann, U. Bovensiepen, R. G.
Moore, J-H. Chu, D. H. Lu, L. Rettig, M. Wolf, I. R. Fisher and Z-X. Shen,
New Journal of Physics \textbf{13}, 063022 (2011).

\bibitem{N1966} B. K. Norling, H. Steinfink, Inorg. Chem. \textbf{5}, 1488
(1966).

\bibitem{FengBilayer2001} D. L. Feng, N. P. Armitage, D. H. Lu, A.
Damascelli, J. P. Hu, P. Bogdanov, A. Lanzara, F. Ronning, K. M. Shen, H.
Eisaki, C. Kim, Z.-X. Shen, J.-i. Shimoyama, and K. Kishio, Phys. Rev. Lett.
\textbf{86}, 5550 (2001).

\bibitem{GarciaNJP2010} David Garcia-Aldea and Sudip Chakravarty, New
Journal of Physics \textbf{12}, 105005 (2010).

\bibitem{HarrisonSciRep2015} N. Harrison, B. J. Ramshaw and A. Shekhter,
Scientific Reports \textbf{5}, 10914 (2015).

\bibitem{SlowYBCO} P.D. Grigoriev and T. Ziman, arXiv:1606.03942.

\bibitem{Commenttb} When SlO originate from the FS warping along $z$-axis,
the angular dependence in Eq. (\ref{Fslow}) has an evident geometrical
interpretation \cite{Yam}. This dependence was also confirmed
quantum-mechanically using the perturbation theory in the first order in the
small parameter $t_{z}/\hbar \omega _{c}\ll 1$,\cite{Kur} and using the
double-layer approach and the Feynman diagram technique.\cite%
{MosesMcKenzie1999,TarasPRB2014} Hence, by analogy one may assume that Eq. (%
\ref{Fslow}) is also valid for the bilayer splitting. However, this analogy
fails in the opposite weak-field regime $t_{b}/\hbar \omega _{c}>1$ when the
SlO appear. The geometrical interpretation similar to Ref. \cite{Yam}, valid
in the weak-field regime $t_{b}/\hbar \omega _{c}>1$, is not applicable for
bilayer splitting or even gives the standard cosine dependence $%
F_{slow}\propto 1/\cos \theta $. Therefore, the problem of the angular
dependence of bilayer splitting at arbitrary $t_{z}/\hbar \omega _{c}$ and $%
\omega _{c}\tau $ needs further theoretical investigation, which is beyond
the scope of this paper.

\bibitem{Commenttz} Eqs. (\ref{tzTheta}) and (\ref{Fslow}) assume the
spatially uniform interlayer hopping, when the interlayer hopping amplitude $%
t_{z}$ does not depend on 2D coordinate within the layer and on in-plane
electron momentum. If the overlaping atomic orbitals are not uniform but
confined within spatial region in the crystalline elementare cell, the
simple dependence in Eqs. (\ref{tzTheta}) and (\ref{Fslow}) may violate, as
e.g. in YBCO high-Tc superconductor.\cite{GarciaNJP2010,Andersen1995}

\bibitem{Banerjee} A. Banerjee, Yejun Feng, D. M. Silevitch, Jiyang Wang, J.
C. Lang, H.-H. Kuo, I. R. Fisher, and T. F. Rosenbaum Phys. Rev. B \textbf{87%
}, 155131 (2013).

\bibitem{CommentElongatedFS} Probably, the pockets of the reconstructed FS
are elongated and oriented along various directions. Their total
contribution to the SlO, being a sum of the contributions from all
individual FS pockets, has a smeared angular dependence of the SlO frequency
$F_{slow}\left( \theta \right) $ as compared to the case of only one
elliptical FS pocket observed in $\beta $-(BEDT-TTF)$_{2}$IBr$_{2}$.\cite{SO}

\bibitem{CommentARPESLowT} Note that the ARPES measurements do not have a
sufficient energy resolution to determine a Fermi-surface reconstruction due
to the second low-$T_{c}$ CDW.\cite{Brouet08,SchmittNJP2011} Therefore, in
ARPES data there is no evidence of such small FS pockets. The transport
measurements, on contrary, are very sensitive to fine FS details, which is
their big advantage as complementary to ARPES technique.

\bibitem{AABook} B.L. Altshuler and A.G. Aronov "Electron-Electron
Interaction In Disordered Conductors", Ch. 1 in "Electron-Electron
Interactions in Disordered Systems", Ed. by A.L. Efros and M. Pollak,
Amsterdam: North-Holland (1985); ISBN: 978-0-444-86916-6.

\bibitem{Kennett2008} Malcolm P. Kennett and Ross H. McKenzie, Phys. Rev. B
\textbf{78}, 024506 (2008).

\bibitem{CommentQHE} A very small interlayer hopping violates the 2D
electron localization in the conducting planes by disorder in magnetic field
\cite{HuckesteinRMP1995}, thus preventing the quantum Hall effect.

\bibitem{HuckesteinRMP1995} B. Huckestein, Rev. Mod. Phys. \textbf{67}, 357
(1995).

\bibitem{CommentManyE} Our study raises several important questions, which
need further experimental and theoretical investigation. For example, the
damping of SlO and of usual MQO amplitudes by the e-e interaction and by
critical fluctuations near an electronic phase transition may strongly
differ. If so, it may serve as an additional tool to measure these
many-particle effects.

\bibitem{GrigorievPRB2008} P.D. Grigoriev, Phys. Rev. B \textbf{77}, 224508
(2008).

\bibitem{ErTe3DoS} R.G. Moore, V. Brouet, R. He, D.H. Lu, N. Ru, J.-H. Chu,
I.R. Fisher, and Z.-X. Shen Phys. Rev. B \textbf{81}, 073102 (2010); F.
Pfuner, P. Lerch, J.-H. Chu, H.-H. Kuo, I.R. Fisher, and L. Degiorgi Phys.
Rev. B \textbf{81}, 195110 (2010).

\bibitem{MosesMcKenzie1999} P. Moses and R.H. McKenzie, Phys. Rev. B \textbf{%
60}, 7998 (1999).

\bibitem{TarasPRB2014} P.D. Grigoriev, T.I. Mogilyuk, Phys. Rev. B \textbf{90%
}, 115138 (2014).

\bibitem{Andersen1995} O. K. Andersen, A.I. Liechtenstein, O. Jepsen, F.
Paulsen, J. Phys. Chem. Solids \textbf{56}, 1573 (1995).
\end{thebibliography}
\end{document}